\documentclass[aps,pre,notitlepage]{revtex4}

\usepackage{amssymb, amsfonts, amsmath}  
\usepackage{graphicx}    % For Graphics

\newcommand{\dir}{Figs}

\begin{document}

%\markboth{F. Schmid}
%{Fluctuations in lipid bilayers}

%\catchline{}{}{}{}{}
%%%%%%%%%%%%%%%%%%%%%%%%%%%%%%%%%%%%%%%%%%%%%%%%%%%%%%%%%%%%%%%%%%%%

\title{FLUCTUATIONS IN LIPID BILAYERS: ARE THEY UNDERSTOOD?}

\author{FRIEDERIKE SCHMID}

\address{Institute of physics, 
Johannes-Gutenberg University of Mainz,
D-55099 Mainz,
Germany\\
Friederike.Schmid@Uni-Mainz.DE}

%\begin{history}
%\received{}
%\revised{}
%\end{history}

\begin{abstract}

We review recent computer simulation studies of undulating lipid bilayers.
Theoretical interpretations of such fluctuating membranes are most
commonly based on generalized Helfrich-type elastic models, with 
additional contributions of local ''protrusions'' 
and/or density fluctuations. Such models provide an excellent 
basis for describing the fluctuations of tensionless bilayers in the
fluid $L_\alpha$ phase at a quantitative level. 

However, this description is found to fail for membranes in the gel phase and
for membranes subject to high tensions. The fluctuations of tilted gel
membranes ($L_{\beta'}$ phase) show a signature of the modulated ripple
structure $P_{\beta'}$, which is a nearby phase observed in the pretransition
regime between the $L_\alpha$ and $L_{\beta'}$ state. This complicates a
quantitative analysis on mesoscopic length scales. In the case of fluid
membranes under tension, the large-wavelength fluctuation modes are found to be
significantly softer than predicted by theory.

In the latter context, we also address the general problem of the relation
between frame tension and the fluctuation tension, which has been discussed
somewhat controversially in recent years. Simulations of very simple model
membranes with fixed area show that the fluctuations should be controlled by
the frame tension, and not by the internal tension.


\end{abstract}

\maketitle

\section{Introduction}

\label{sec:introduction}

Lipid bilayers in aqueous solution have been studied intensely for many decades
as simple model systems for biological
membranes\cite{Gennis89,BEM91,Alberts02,TN04}, both
experimentally\cite{KC94,KC98,KC02} and by computer simulations (see Refs.\
\onlinecite{VSK06,MKS06,MVT08,Heimburg07,Brown08,Deserno09,Review09,Stecki10,LR10}
for recent reviews).  The physical and thermodynamical properties of such
bilayers are similar in many respect for a broad variety of
lipids\cite{BEM91,KC94,KC98,KC02}. 
For example, phase diagrams of single-component lipid bilayers have a
generic structure.  At high temperatures, the bilayers are in a ''fluid'' state
($L_{\alpha}$ phase) where the hydrophobic tails of the lipids are disordered
and the lipids are mobile.  At lower temperatures, they undergo a transition
into a stiffer and more ordered ''gel'' phase with ordered tails, reduced lipid
mobility, and hexatic positional order.  The detailed structure of that state
depends on the type of the hydrophilic head groups, and most notably, on their
size compared to the hydrophobic tails.  If the heads are small compared to the
tails -- {\em e.g.}, for ethanolamines\cite{KC94} -- the tails point on average
in the direction of the membrane normal ($L_\beta$ phase). If the heads are
large -- {\em e.g.}, in the common class of phosphatidylcholines\cite{KC98} --
the chains adjust to the packing mismatch by collective tilting ($L_{\beta
'}$ phase). In some cases where the coupling between them is weak, {\em e.g.},
ether-linked phosphatidylcholines\cite{KC98}, they may even give up the bilayer
structure and assemble into a straight interdigitated phase ($L_{\beta
\mbox{\tiny INT}}$ phase). The transition between the $L_\alpha$ phase and the
$L_{\beta'}$ phase or the $L_{\beta \mbox{\tiny INT}}$ phase proceeds via an
intermediate modulated ``ripple'' phase ($P_{\beta'}$).  In nature, most
biomembranes seem to be maintained in a fluid state.  Nevertheless, it is
remarkable that the transition between the fluid and the gel state, the
so-called main transition, occurs at temperatures well above room temperature
for some of the most common lipid molecules ({\em e.g.}, $\sim 41{}^0$C for the
phospholipid dipalmitoylphosphatidyncholine (DPPC)\cite{KC98}). One can
speculate that nature might use the vicinity of such a phase transition to 
tune central properties of the membrane such as its 
permeability\cite{GWH10,Heimburg10}.

One striking property of lipid bilayers is their softness and the ensuing large
thermal undulations. Membrane fluctuations have attracted persistent interest
over the years, going back to a seminal paper by Helfrich\cite{Helfrich73} in
1973 up to today. They affect the interactions of membranes with other objects,
{\em e.g.}, in the context of adhesion, lipid-mediated protein interactions and
lateral protein diffusion. Moreover, the detailed analysis of fluctuations
allows one to extract elastic constants and establish a connection between
experiments or molecular models and mesoscale continuum theories of membranes.
Such an analysis requires a detailed knowledge of the relation between the
molecular fluctuations and the membrane elasticity, and it thus relies strongly
on excellent theoretical descriptions.

In the present paper, we review simulation results for fluctuating lipid
bilayers, with a strong focus on own work. We first give a very brief and
incomplete overview over the history and the current state of membrane
fluctuation theories for planar membranes. Then, in Section \ref{sec:general},
we focus on fluid membranes, construct a general form of membrane Hamiltonian
for coupled monolayers and address in particular the issue of membrane tension.
In Section \ref{sec:model}, we discuss own simulation results with a generic
coarse-grained simulation model. We have studied fluctuating membranes in the
liquid phase and the gel phase, and also under tension. The theoretical
description is far from satisfactory in many respect, and further refinements
are necessary.

\section{Fluctuations of flat membranes: A brief overview}

\label{sec:overview}

The oldest and simplest Ansatz for the curvature elastic free energy density
per unit area of a fluid membrane is the famous Helfrich
Hamiltonian\cite{Helfrich73} 
\begin{equation}
\label{eq:helfrich0}
  {\cal H} = \int {\rm d}A \: \Big\{ \frac{k_c}{2} (H - 2c_0)^2 
      + \bar{\kappa} K \Big\}.
\end{equation}
Here $\int {\rm d}A$ is a surface integral running over the membrane area $A$,
$H$ the total curvature at a given membrane
position, {\em i.e.}, the sum of the inverse curvature radii, and $K$ is the
Gaussian curvature, {\em i.e.}, their product. The elastic parameters
$k_c$, $\bar{\kappa}$, and $c_0$ are the bending rigidity, the Gaussian
rigidity, and the spontaneous curvature of the membrane.  For symmetric
bilayers, $c_0$ is zero. The integral over $K$ on closed surfaces
contributes a constant which only depends on the topology, according to the
Gauss-Bonnet theorem. Hence it can be omitted if the topology is fixed. 
Deuling and Helfrich later introduced a generalized Hamiltonian which 
includes a ''surface tension'' term $\Gamma \int {\rm d}A$ that couples 
directly to the membrane surface in a Lagrange multiplier sense\cite{Deuling76}
\begin{equation}
\label{eq:helfrich}
  {\cal H} = \int {\rm d}A \: \Big\{ \Gamma + \frac{k_c}{2} (H - 2c_0)^2 
      + \bar{\kappa} K \Big\}.
\end{equation}

From Eq.\ (\ref{eq:helfrich}), one can derive that flat fluid membranes at
$\Gamma=0$ should have soft long-wavelength modes that diverge as $q^{-4}$ with
the wavevector $q$. With few exceptions (see section \ref{sec:general} for a
more detailed discussion), most researchers have assumed that such membranes
must be ''tensionless'' in the sense that they do not experience mechanical
stress \cite{GL98,GSL09}.  Experimentally, this is not a common situation: When
membranes form closed vesicles, tension usually builds up due to osmotic
pressure differences between the inside and the outside of the vesicle.
Tension also arises in suspended membranes clamped to a flat frame. Thus the
''tensionless'' case is mostly of theoretical interest.  In computer
simulations, it can be implemented in a straightforward manner by employing
simulation methods where the shape of the simulation box is allowed to
fluctuate\cite{FS96}, or by varying the system dimensions until the surface
tension as evaluated from the integrated stress profile across the membrane
vanishes\cite{GL98}. The $q^{-4}$-spectrum was first verified in computer
simulations of a coarse-grained bilayer model by Goetz, Gompper, and
Lipowsky\cite{GGL99}, and later confirmed by numerous later studies of
coarse-grained as well as atomistic
models\cite{GGL99,LE00,MM01,LMK03,WF05,BB06,WBS09,BBS11}.  Coupled Helfrich
Hamiltonians were also found to provide a suitable framework for a quantitative
theoretical description of fluctuating membrane
stacks\cite{L95,LMK03,LMM04,LMM05}.

While the Helfrich Hamiltonian captures nicely the long-wavelength fluctuations
of membranes, it is not designed for describing the structure and fluctuations
on small wavelengths, {\em i.e.}, of the order of the membrane thickness.
Several extensions have been proposed to improve the description of membranes
on these molecular scales within continuum approaches. Lipowsky and coworkers
\cite{LG93,LG93b,GGL99} introduced the notion of ''protrusions'', {\em i.e.},
independent fluctuations of single lipids that supposedly govern the
fluctuation spectrum of membranes on the molecular level. Recently, Brandt et
al.\cite{BBS11,BBE11} suggested that these protrusions correspond to lipid
density fluctuations within the membrane, and corroborated this picture with
atomistic and coarse-grained simulations (see also Section \ref{sec:fluid}).
One way to analyze the molecular scale fluctuations of membranes in more
detail, pioneered by Lindahl and Edholm, is to consider separately the height
and thickness fluctuations\cite{LE00}. Thickness distortions have traditionally
played a role in elastic theories for membrane mediated interactions between
inclusions\cite{DPS93,DBP94,ABD96}.  Extending these approaches to the
fluctuation problem, Brannigan and Brown have proposed a continuum model which
allows to study height and thickness fluctuations and relate them to
protein-protein interactions in one unified framework\cite{BB06,BB07}. The
Helfrich Hamiltonian has also been extended to include internal degrees of
freedom, such as local tilt\cite{MNK07,WPW11}. 

Most studies in the past have focused on fluid membranes. In the gel phase, the
situation is complicated by the fact that the lipids have positional order.
Already in 1987, Nelson and Peliti\cite{NP87} have analyzed by analytical
theory the fluctuations of crystalline and hexatic membranes, which can sustain
lateral elastic shear stress to some extent. They argued that crystalline
membranes should stiffen due to fluctuations on larger length scales, such that
the renormalized effective bending rigidity becomes $q$-dependent and scales as
$k_c(q) \propto 1/q$. Their calculation was later refined by Le Doussal
and Radzihovsky\cite{DR92}, who obtained $k_c(q) \propto q^{-0.821}$
within a self-consistent screening approximation. For hexatic membranes, the
strict positional order is destroyed due to the presence of free dislocations,
but lipids still have long range 'bond-orientational' order, {\em i.e.}, the
vectors connecting nearest neighbor lipids have well-defined average
orientations. According to Nelson and Peliti, the bending stiffness then still
increases for large length scales, but only logarithmically, $k_c(q) \sim
\sqrt{- \ln(q \xi_T)}$, where $\xi_T$ is a translational correlation length.
Another factor that must be considered in the gel $L_{\beta'}$ phase is the
collective tilt of the molecules.  According to Park, however, the qualitative
behavior of the bending rigidity does not differ for tilted\cite{Park96} and
untilted\cite{PL95} hexatic membranes in the so-called ''strong coupling
limit''. One might note that the bending rigidity should also slightly be
renormalized in fluid membranes by fluctuations - in this case, they soften
them\cite{PL85}. However, this only becomes relevant on length scales where the
membranes exhibit large deviations from planar.

\section{General considerations for fluid membranes}

\label{sec:general}

Returning to fluid membranes, we will now present a general framework for
studying fluctuations of simple fluid membranes (without internal degrees of
freedom) within a continuum theory. Following Brannigan and Brown\cite{BB06}
and their predecessors\cite{LG93b,DPS93,DBP94,ABD96}, we describe planar membranes by
a system of two coupled elastic surfaces, which correspond to the interfaces
between the lipid region and the surrounding aqueous solvent. The membranes are
taken to be almost flat and have no bubbles or overhangs, thus their surfaces
can be parameterized by two unique functions $z_{1,2}(x,y)$ of the projected
coordinates $x$ and $y$.  Equivalently, they can be characterized by their
local height $H(x,y) = (z_1(x,y) + z_2(x,y))/2$ and thickness $2 T(x,y) =
(z_1(x,y) - z_2(x,y))$ and for simplicity, the coordinate system is chosen such
that the mean height $\langle H \rangle$ is zero. We expand the Hamiltonian up
to second order in the small parameters $H$ and $u = T - t_0$, where $2 t_0$ is
the equilibrium thickness of the membrane, keeping spatial derivatives up to
second order.  Based on a few elementary symmetry assumptions,
\begin{itemize}
\item[(i)] Isotropy within the membrane plane
\item[(ii)] Translational invariance within the membrane and in the $z$ 
   direction
\item[(iii)] Invariance with respect to flipping $H \to - H$ ({\em i.e.},
the bilayer is symmetric.)
\end{itemize}
we can then construct the following very general Hamiltonian for the
bilayer fluctuations of a membrane with projected area $A_p$.
\begin{eqnarray}
\nonumber
 {\cal H} &=& \frac{1}{2} \int_{A_p} \: {\rm d}x \: {\rm d}y \: \Big\{
  a_2 (\nabla H)^2 + a_3 (\Delta H)^2
   + b_1 u^2 + b_2 (\nabla u)^2 + b_3 (\Delta u)^2 \Big\}
   \\ && + \; \mbox{boundary terms}.
\label{eq:h_general}
\end{eqnarray}
Other terms are either forbidden by symmetry (such as all crossterms containing
both $H$ and $u$) or can be incorporated into the boundary terms. For example,
the curvature-like term $\Delta u$ gives a boundary term, and $u \Delta u$ can
be converted into $(\nabla u)^2$ (plus boundary terms) by means of a partial
integration. 

We emphasize that the derivation of Eq.\ (\ref{eq:h_general}) does not rely on
any specific assumptions regarding the underlying physics. For example, nowhere
was it necessary to require that we are dealing with bilayers consisting of
well-separated monolayers, that lipids have fixed volume, etc.  These
assumptions come in when finding an {\em interpretation} for the elastic
parameters $a_i$ and $b_i$.  Brannigan and Brown\cite{BB06} have derived a
Hamiltonian of the form (\ref{eq:h_general}) based on a physical picture where
${\cal H}$ describes the bending modes of two coupled monolayers, each with
area compressibility, spontaneous curvature, and bending rigidity, and subject
to the constraint that the volume per lipid is constant.  In this case, the
tensionlike parameter $a_2= \Gamma$ vanishes for tensionless bilayers, the
parameter $b_1$ can be associated with the area compressibility $k_A$ via $b_1
= k_A/t_0^2$, the parameters $a_3$ and $b_3$ are identified with the bending
rigidity $k_c$ of the bilayer, $a_3 = b_3= k_c$, and the parameter $b_2$ is
related to the spontaneous curvature $c_0$ of monolayers, $b_2 = - 4 k_c
\zeta/t_0$, where $\zeta = c_0 - \Sigma {\rm d}c_0/{\rm d}\Sigma$ depends on
$c_0$ and its derivative with respect to the lipid area, $\Sigma$. In addition,
Brannigan and Brown introduced additional protrusion fields 
$\lambda_{1,2}(x,y)$ -- following Lipowsky and Grotehans \cite{LG93b} --
which describe short-wavelength fluctuations on both
surfaces and are taken to be independent of each other and of the bending
degrees of freedom $z_i(x,y)$. The total height and thickness are thus given by
$h(x,y) = H + (\lambda_1+\lambda_2)/2$ and $t(x,y) = T +
(\lambda_1-\lambda_2)/2$, and the fluctuation spectrum in Fourier space
reads\cite{BB06,NWN10}

\begin{eqnarray}
 \label{eq:elastic_height} 
  \langle | h_{\mathbf{q}} |^2 \rangle &=& \frac{k_B T}{\Gamma q^2 + k_c q^4}
      + \frac{k_B T}{2(k_{\lambda} + \gamma_{\lambda}q^2)}\\
\label{eq:elastic_width} \langle | t_{\mathbf{q}} |^2 \rangle &=&
 \frac{k_B T}{k_A/t_0^2 - 4 k_c \zeta q^2 / t_0 + k_c q^4 } +
 \frac{k_B T}{2(k_{\lambda} + \gamma_{\lambda}q^2)} \quad ,
\end{eqnarray}
where the parameters $\gamma_\lambda$ and $k_\lambda$ 
characterize the protrusion modes.

The parameter $\Gamma$ deserves special attention\cite{NWN10}. It depends on
the ''frame tension'' $\Gamma_{\mbox{\tiny frame}}$, {\em i.e.}, the lateral
stress within the projected membrane plane (the ($x,y$) plane). The frame
tension is a mechanical tension which serves to maintain the projected area
$A_p$ (corresponding to the ''lateral tension'' in Ref. \cite{GL98}).
The parameter $\Gamma$ is an intrinsic coefficient characterizing the
membrane fluctuations.  Therefore, the relation between $\Gamma$ and
$\Gamma_{\mbox{\tiny frame}}$ is not clear {\em a priori}.

The problem already arises when considering single fluctuating surfaces (no
protrusions) governed by the Helfrich Hamiltonian (\ref{eq:helfrich}). For
that case, it has been discussed intensely and somewhat controversially in the
past\cite{BGP76,DL91,CLN94,Jaehnig96,Marsh97,FP03,FP04,Imparato05,Imparato06,Stecki08,FB08,Schmid11,Farago11,Haim11,Schmid12}.
For single almost flat surfaces, the linearized Helfrich Hamiltonian
reads\cite{Safran94}
\begin{equation}
\label{eq:helfrich_plane}
 {\cal H} = \frac{1}{2} \int_{A_p} \: {\rm d}x \: {\rm d}y \; \Big\{
  \Gamma (\nabla h)^2 +  k_c (\Delta h)^2 \Big\},
\end{equation}
(the contribution of the Gaussian curvature has been omitted), and the
corresponding fluctuation spectrum in Fourier space takes the form 
\begin{equation}
\label{eq:helfrich_fluct}
\langle |
h_{\mathbf{q}} |^2 \rangle = \frac{k_B T}{ \Gamma q^2 + k_c q^4}.
\end{equation}
Assuming that the number of degrees of freedom (the number of lipids) is fixed, 
the free energy $F$ of the linearized Hamiltonian (\ref{eq:helfrich_plane}) can be
evaluated exactly as a function of $\Gamma$ and the projected area $A_p$. The
frame tension can then be evaluated as the derivative $\Gamma_{\mbox{\tiny
frame}} = \partial F/\partial A_p$, giving\cite{FP03}
\begin{equation}
\Gamma_{\mbox{\tiny frame}} = \Gamma \left( 1 + \frac{k_B T}{8 \pi k_c} \ln
\Big( 1 + \frac{4 \pi k_c N}{\Gamma A_p} \Big) \right)
- \frac{k_B T \: N}{2 A_p},
\end{equation}
which clearly differs from $\Gamma$. Based on this observation, some 
authors\cite{Imparato06,Stecki08,FB08} have recently argued that the tensionlike
parameter in the fluctuation spectrum should differ from the frame tension -- more
precisely, they claimed that the ''fluctuation tension'' $\Gamma_{\mbox{\tiny
fluc}}$, defined as the coefficient which drives the $q^2$-contribution to the
inverse fluctuation spectrum in an equation of the form (cf. Eq.~(\ref{eq:helfrich_fluct}))
\begin{equation}
\langle | h_{\mathbf{q}} |^2 \rangle = \frac{k_B T}{\Gamma_{\mbox{\tiny fluc}} q^2 + {\cal O}(q^4)}
\end{equation}
differs from the frame tension $\Gamma_{\mbox{\tiny frame}}$. Within the linear
theory (\ref{eq:helfrich_plane}), one of course has $\Gamma_{\mbox{\tiny fluc}} = \Gamma$,
but this is not true in general. 

The problem with this argument is that the Hamiltonian
(\ref{eq:helfrich_plane}) is not exact -- it is just the linearized version of
Eq.\ (\ref{eq:helfrich}).  Nonlinear contributions renormalize
$\Gamma_{\mbox{\tiny frame}}$, and, even more importantly, $\Gamma_{\mbox{\tiny
fluc}}$. Cai {\em et al.}\cite{CLN94} have examined the fluctuations of
membranes fixed area per lipid in the grand canonical ensemble $(\mu,A_p,T)$,
where the number of lipids is allowed to fluctuate. (In this case the frame
tension is related to the grandcanonical potential {\em via} $\Omega(\mu,A_p,T)
= \Gamma_{\mbox{\tiny frame}} A_p$.) They argued on very general grounds that
the requirement of gauge invariance -- invariance with respect to a rotation of
the ''projected plane'' -- implies $\Gamma_{\mbox{\tiny frame}} =
\Gamma_{\mbox{\tiny fluc}}$ in the thermodynamic limit. Farago and
Pincus\cite{FP04,Farago11} presented a similar argument for the case of
membranes with fixed projected area and fixed number of lipids ($(N,A_p,T)$
ensemble).  

\begin{figure}
  \centering
  \includegraphics[width=0.7\textwidth]{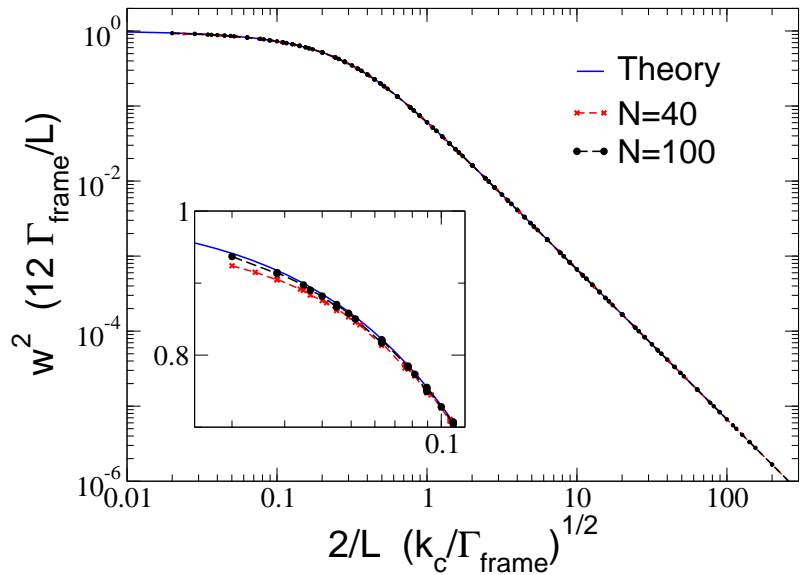} \\
  \caption{Squared amplitude of fluctuations $w^2$ of flat one-dimensional
membranes (stiff lines) vs. bending rigidity $k_c$, rescaled in a manner
suggested by the linearized elastic theory (\protect\ref{eq:helfrich_plane}),
where the fluctuation tension is taken to be the frame tension $\Gamma_{\mbox{\tiny frame}}$ and
the ''projected length'' of the membrane $L_p$ is replaced by the actual length
$L$.  Data have been collected for $k_c/L = 1-1000$ and $\Gamma_{\mbox{\tiny frame}} L =
0.05-10000$ and two values $N$ of discretization along $L$. Solid line gives
the scaling function predicted by theory. Inset shows blowup of a region that
emphasizes the effects of finite discretization. }
  \label{fig:width_1d}
\end{figure}

Now the existence of the thermodynamic limit must be questioned for ''flat''
membranes in the floppy low-tension regime, because they bend around on length
scales larger than the persistence length. Therefore, infinitely large
membranes at low tension are no longer planar. Furthermore, infinitely large
membranes at finite tension can always lower their free energy by forming a
pore, therefore they are not at thermodynamic equilibrium in the
''thermodynamic limit''\cite{GSL09}.  If there is no thermodynamic limit,
different ensembles are no longer equivalent, and results for the $(\mu,A_p,T)$
ensemble or the $(N,A_p,T)$ ensemble are not necessarily valid for the
$(N,\Gamma_{\mbox{\tiny frame}},T)$ ensemble, which is practically more
relevant and often implemented in simulations.  To clarify the situation, the
present author has carried out high-precision computer simulations of a
one-dimensional infinitely thin ''membrane'', {\em i.e.}, a stiff
one-dimensional line, in the $(N,\Gamma_{\mbox{\tiny frame}},T)$ ensemble for
frame tensions $\Gamma_{\mbox{\tiny frame}}$ and bending rigidities $k_c$
spanning several orders of magnitude\cite{Schmid11}. It turned out that the
fluctuations could be described very accurately by analytical predictions of
the linearized theory (\ref{eq:helfrich_plane}), {\em if} the tension parameter
$\Gamma$ is replaced by a renormalized tension parameter $\Gamma_{\mbox{\tiny
frame}}$.  This is demonstrated in Fig.\ \ref{fig:width_1d} for the example of
the squared amplitude of fluctuations $w^2 = \langle h^2 \rangle - \langle h
\rangle^2$.

The numerical results thus show that the fluctuations are indeed driven by the
frame tension. They were confirmed by subsequent simulations of Farago,
also of one dimensional membranes, which emphasize the role of the gauge
invariance\cite{Farago11}: For Hamiltonians which are not gauge invariant,
$\Gamma_{\mbox{\tiny frame}} = \Gamma_{\mbox{\tiny fluc}}$ is no longer valid.

Summarizing and concluding, Eq.\ (\ref{eq:helfrich_plane}) can be used to
calculate fluctuation spectra of single membranes, if it is interpreted as an
effective Hamiltonian with renormalized coefficients, $\Gamma \to
\Gamma_{\mbox{\tiny fluc}}$ and $k_c \to k_{c,\mbox{\tiny fluc}}$.  In
particular, the fluctuation tension renormalizes to the frame tension,
$\Gamma_{\mbox{\tiny fluc}} = \Gamma_{\mbox{\tiny frame}}$. The bending
rigidity should be renormalized as well\cite{PL85,FP04,GLK06}, but in practice, the
renormalization effect is negligeable (as mentioned in Sec.\
\ref{sec:overview}, $k_c$ softens slightly on large length scales). This
conclusion should also hold for more sophisticated elastic membrane models such
as the coupled surface models, Eqs.\ (\ref{eq:h_general}) and
(\ref{eq:elastic_height}). Thus we conclude that the parameter $\Gamma$ should
be identified with the frame tension.

\section{Fluctuations in a generic coarse-grained model for lipid bilayers}
 \label{sec:data}

The general theoretical considerations presented in the previous sections can best
be tested by comparison with explicit molecular simulations of bilayers. In
this section, we discuss a series of simulations of bilayers in different states,
fluid and gel, and under tension, in a simple lipid bilayer
model\cite{LS05,LS07,SDL07}. This model has been constructed in the spirit of
coarse-grained approaches\cite{VSK06,Review09} from a successful model for Langmuir
monolayers\cite{HHB95,HHB96,HH96,SLS99,SS99,DS01} and has been shown to reproduce the
most important membrane phases and phase transitions of DPPC\cite{LS07}, {\em i.e.}, 
the transitions between the $L_\alpha$-phase, the modulated $P_{\beta'}$ phase, and
the $L_{\beta'}$-phase. It has also been used to test predictions of
the elastic theory regarding membrane-protein and membrane-mediated
protein-protein interactions\cite{WBS09,NWN11}. Here, we will focus on
the undulations.

\subsection{The Model}

\label{sec:model}

In our model\cite{SDL07,LS07}, each ''lipid'' consists of a chain of seven
beads with one head bead of diameter $\sigma_h$ followed by six tail beads of
diameter $\sigma_t$.  Non-bonded beads interact, {\em via} a truncated and
lifted Lennard-Jones potential:
\begin{equation}
  \label{eq:lj_bead}
  V_{\mbox{\tiny bead}}(r) = \left \{
  \begin{array}{rl}
    V_{\mbox{\tiny LJ}}(r/\sigma) - V_{\mbox{\tiny LJ}}(r_c/\sigma) 
            & \mbox{  if $r < r_c$} \\
          0 & \mbox{  otherwise}
\end{array}
\right.
\end{equation}
with
\begin{equation}
  \label{eq:lj}
  V_{\mbox{\tiny LJ}}(x) = \epsilon \left( x^{-12} - 2 x^{-6} \right)
\end{equation}
where $\sigma$ is the mean diameter of the two interacting beads, $\sigma_{ij}
= (\sigma_i + \sigma_j)/2$ ($i,j = h$ or $t$).  Head-head and head-tail
interactions are purely repulsive ($r_c = \sigma$), while tail-tail
interactions also have an attractive contribution ($r_c = 2\sigma$). Bonded
beads are connected by FENE (Finitely Extensible Nonlinear Elastic) springs
with the spring potential
\begin{equation}
  V_{\mbox{\tiny FENE}} (r) = -\frac{1}{2} \epsilon_{_{\mbox{\tiny FENE}}}
     (\Delta r_{\mbox{\tiny max}})^2
  \log \left (1 - \left ( \frac{r - r_0}{\Delta r_{\mbox{\tiny max}}} \right )^2
  \right ),
  \label{eq:fene}
\end{equation}
where $r_0$ is the equilibrium distance, $\Delta r_{\mbox{\tiny max}}$ the
maximal deviation, and $\epsilon_{\mbox{\tiny FENE}}$ the FENE spring constant.
In addition, a bond-angle potential applies,
\begin{equation}
  \label{eq:ba}
  V_{BA}(\theta) = \epsilon_{BA} (1 - \cos(\theta)).
\end{equation}
The aqueous environment of the membrane is modeled with ``phantom'' solvent
beads\cite{LS05}, which interact with lipids like head beads ($\sigma_s =
\sigma_h$), but have no interactions with each other.

Systems of up to $7200$ lipids were studied using Monte Carlo simulations at
constant pressure, temperature, and surface tension with periodic boundary
conditions.  Here the pressure couples to the total volume $V$ of the
simulation box, and the surface tension couples to the total area $A$.  The
simulation box is a parallelepiped and all side lengths and angles fluctuate
independently during the simulation.  Run lengths were up to to $8$ million Monte
Carlo steps, where one Monte Carlo step corresponds to one Monte Carlo move per
bead, and moves that change the shape of the simulation box were attempted
every 50th Monte Carlo step. The code was parallelized using a domain
decomposition scheme described in Ref.~\onlinecite{SDL07}. To study the
fluctuations of the membrane, we have determined the mean head positions
$z_1(x,y)$, $z_2(x,y)$ of the two monolayers on a grid $(x,y)$, and derived the
height and thickness profiles $h(x,y) = (z_1 + z_2)/2$ and $t(x,y) = (z_1 -
z_2)/2$.  The spectra were Fourier transformed according to
\begin{equation}
f_{q_x,q_y} = \frac{L_x L_y}{N_x N_y} \sum_{x,y} f(x,y)
e^{-i (q_x x + q_y y)},
\end{equation}
and averages $\langle |h_q |^2 \rangle$ and $\langle |t_q |^2 \rangle$ were
evaluated in $q$-bins of size $0.1/\sigma_t$.  The procedure is described in
more detail in Refs.~\onlinecite{LMK03,WBS09}.

\begin{figure}
  \centering
  \includegraphics[width=0.6\textwidth]{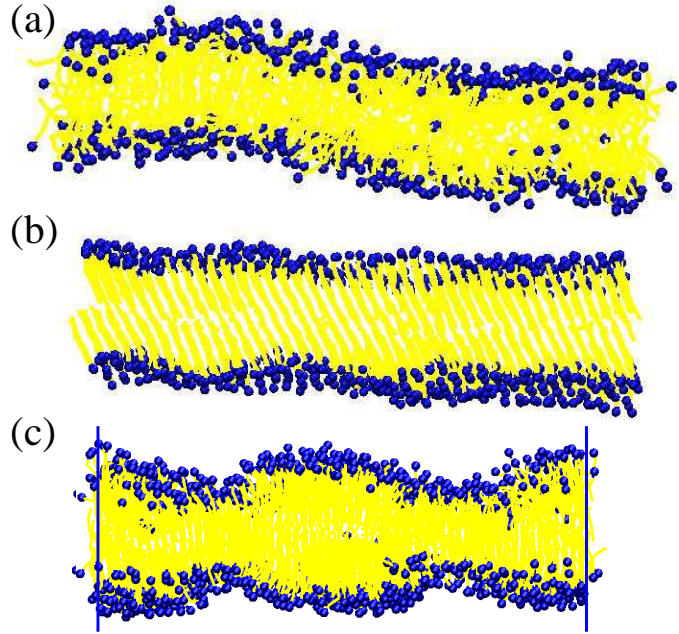} \\
  \caption{Slices through configurations of our model lipid bilayers 
  (a) in the fluid phase $L_\alpha$ ($T=1.3\epsilon/k_B$),
  (b) the gel phase $L_{\beta'}$ ($T=1 \epsilon/k_B$), and 
  (c) the asymmetric ripple phase $P_{\beta'}$ ($T=1.18 \epsilon/k_B$)
  }
  \label{fig:snapshots}
\end{figure}

The model parameters are\cite{DS01,SDL07} $\sigma_h = 1.1 \sigma_t$, $r_0 =
0.7\sigma_t$, $\Delta r_{\mbox{\tiny max}} = 0.2 \sigma_t$,
$\epsilon_{_{\mbox{\tiny FENE}}} = 100 \epsilon/\sigma_t^2$, and $\epsilon_{BA}
= 4.7 \epsilon$, and the pressure was chosen $P = 2. \epsilon/\sigma_t^3$. At
these parameters, the lipids spontaneously self-assemble into stable bilayers,
the stable low-temperature phase is the tilted gel phase
$L_{\beta'}$, and an intermediate rippled $P_{\beta'}$ state emerges in the
transition region between the fluid an the gel state. The main transition
occurs at the temperature\cite{LS07} $k_B T = 1.2 \epsilon$. By reducing the
size $\sigma_h$ of the head beads, one can suppress the tilt in the gel state,
such that the lipids assume the straight $L_\beta$ structure. In this case the
ripple phase also disappears\cite{Dolezel_Thesis}, which underlines the
generic relation between ripple and tilt. In the present paper, however,
we focus on membranes that do exhibit tilt at low temperatures 
(see Fig. \ref{fig:snapshots}).

To map the model units onto standard SI units, we first compare the bilayer
thickness ($2 t_0 \sim 6 \sigma_t$ in the $L_{\alpha}$ phase and 
$2 t_0 \sim 7.7 \sigma_t$ in the $L_{\beta'}$ phase) and the area per lipid 
($a \sim 1.4 \sigma_t^2$ in the $L_{\alpha}$ phase and 
$a \sim 1 \sigma_t^2$ in the $L_{\beta'}$ phase) with the
corresponding numbers for real lipid bilayers. The ratios $a/t_0^2$ in our
model bilayers roughly matches that of DPPC bilayers in the respective
phases, thus we identify our model lipids with DPPC molecules. 
Either $t_0$ or $a$ is then used to estimate the length scale, giving\cite{Lenz_Thesis}
$\sigma_t \sim 6$\AA. The energy scale can be estimated 
by matching the temperatures of the main transition, giving 
$\epsilon \sim 0.36 \cdot 10^{-20}$J.

\subsection{Fluid phase}

\label{sec:fluid}

\begin{figure}
  \centering
  \includegraphics[width=0.6\textwidth]{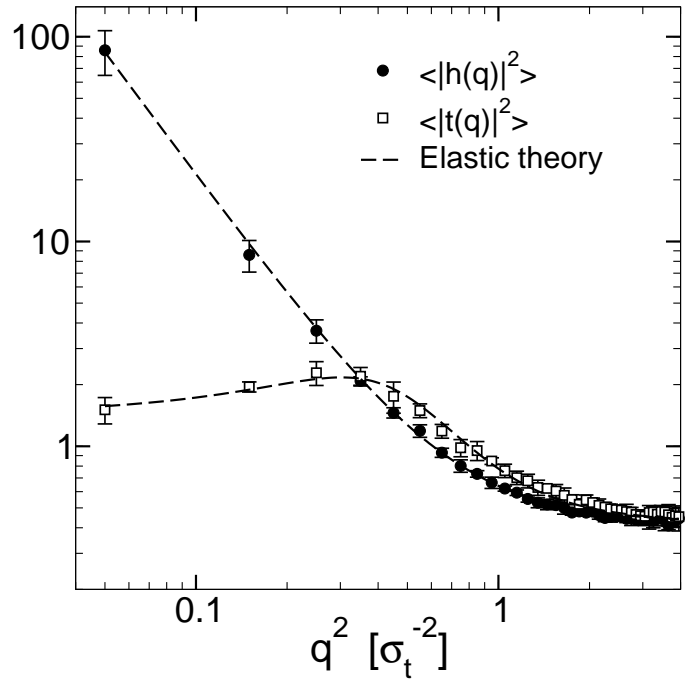}\\
  \caption{Radially averaged Fourier spectrum of the height (full circles)
  and thickness (open squares) fluctuations of tensionless membranes in
  the fluid state ($T=1.3 \epsilon$). Dashed lines show the best fit to 
  the elastic theory, Eq.\ (\ref{eq:elastic_height}) and (\ref{eq:elastic_width}).
  After Ref.\ \protect\onlinecite{WBS09}.
  }
  \label{fig:fluc_fluid}
\end{figure}

We begin with discussing the tensionless fluid state of membranes, which has
received by far the most attention in the past. Fluctuations of tensionless
fluid membranes have been studied by numerous authors using
atomistic\cite{LE00,MM01} as well as
coarse-grained\cite{GGL99,CD05,WF05,BB06,Imparato06,Stecki08,WBS09,BBS11}
models.  With a few exceptions\cite{Imparato06,Stecki08}, most studies recover
the expected $q^{-4}$ divergence of the height fluctuations at low wavevectors
$q$. At high $q$, the height fluctuations are dominated by protrusions and
level off.  Fig.\ \ref{fig:fluc_fluid} shows the results of the fluctuation
analysis for our model\cite{WBS09}, which also includes an analysis of the
thickness fluctuations.  Since membranes have a global equilibrium thickness,
the thickness fluctuations stay finite at low $q$. With increasing $q$, they
rise and exhibit a broad maximum around $q^2 \sim 3 \sigma_t^{-2}$,
corresponding to a soft peristaltic mode with wavelength around $10 \sigma_t$.
At high $q$, the thickness fluctuation spectrum follows closely the height
fluctuation spectrum\cite{BB06}.  This is consistent with Eqs.\
(\ref{eq:elastic_height}) and (\ref{eq:elastic_width}) and a direct consequence
of the fact that the protrusion fluctuations of the two monolayers are
independent of each other and of the bending fluctuations. Our simulation data
can be fitted very nicely with the prediction of the elastic theory, Eqs.\
(\ref{eq:elastic_height}) and (\ref{eq:elastic_width}) with $\Gamma = 0$. 

An interesting alternative way of analyzing the data has very recently been
introduced by Brandt et al.\cite{BBS11}. Instead of determining continuous
surfaces $z_{1,2}(x,y)$ in real space by a binning procedure and then Fourier
transforming these functions, they proposed to analyze directly the Fourier
transform of the molecular positions. i.e., the quantities
\begin{equation}
\bar{z}_\alpha({\bf q}) = 
\frac{1}{N_\alpha} \sum_{l=1}^{N_\alpha} (z_l-\langle z \rangle) \:
{\rm e}^{i {\bf q}\cdot {\bf r}_l},
\end{equation}
where the sum $l$ runs over the $N_\alpha$ lipids in monolayer $\alpha$
($\alpha=1,2$), ${\bf r}_l$ gives the positions of the head group of these
lipids, and $\langle z \rangle = \frac{1}{N_\alpha} \sum_1^{N_{\alpha}} z_l$
gives the average height of the head groups within a monolayer.  In the spirit
of Section \ref{sec:general}, one can define a height and a thickness spectrum
from the quantities $\bar{h}=\frac{1}{2}(\bar{z}_1+\bar{z}_2)$ and
$\bar{u}=\frac{1}{2}(\bar{z}_1-\bar{z}_2)$. On molecular length scales ($q > 1
{\rm nm}^{-1}$), the two spectra are equal and proportional to the number
density structure factor, $\langle |\bar{h}({\bf q})|^2 \rangle = \langle
|\bar{u}({\bf q})|^2 \rangle = C a^2 \langle | \rho({\bf q})|^2 \rangle$, where
the proportionality factor $C = (\sigma_{z,1}^2 + \sigma_{z,2}^2)/2$ is the
mean width of the distributions of $z$-positions within the monolayers 1 and 2.
After subtracting the contribution of density fluctuations, the remaining
height spectrum $(\langle |\bar{h}|^2 \rangle - C a^2 \langle | \rho({\bf
q})|^2 \rangle)$ was found to follow closely a $q^{-4}$ behavior down to the
noise level, and on length scales even beyond $q \sim 1 {\rm nm}^{-1}$. This
was observed consistently both in coarse-grained and atomistic
simulations\cite{BBS11,BBE11}, and it strongly suggests that the
''protrusions'' observed in the binning method are really a signature of
density fluctuations. 

The ''direct Fourier method'' by Brandt et al.\cite{BBS11} provides new
insights and is more elegant than the binning method, because it does not
involve the somewhat arbitrary binning procedure. However, the comparison of
the resulting spectra with theory is less straightforward than in the binning
method, since the quantity $\bar{z}$ is the Fourier transform of a product,
$z(x,y) \: \rho(x,y)$.  In the remainder of the paper, we will only discuss
results obtained with the binning method.

Coming back to our model, the fit to the elastic theory shown in Fig.\
\ref{fig:fluc_fluid} allows us to extract the elastic constants of our model
membrane, most notably the bending stiffness $k_c = 6.2\pm 0.4 \epsilon \sim
2.2 \cdot 10^{-20} {\rm J}$ and the rescaled area compressibility $k_A/t_0^2 =
1.3 \pm 0.3 \epsilon/\sigma_t^4 \sim 3.6 \cdot 10^{-20} {\rm J/nm}^4$.  As
another elastic parameter, the spontaneous curvature of a monolayer can be
obtained from the first moment of the pressure profile across the
monolayer\cite{Safran94}, giving\cite{WBS09} $c_0 = -0.05 \pm 0.02
\sigma_t^{-1} \sim -0.08 {\rm nm}^{-1}$. These elastic constants have the same
order of magnitude than those obtained from all-atom simulations of
DPPC\cite{LE00}, $k_c \sim 4 \cdot 10^{-20}$J, $k_A/t_0^2 \sim 1.1 \cdot
10^{-20}$J/nm${}^4$, and\cite{MRY07} $c_0 \sim -0.04$ -- $-0.05$ nm${}^{-1}$,
or from experimental estimates\cite{Marsh06}, $k_c \sim 5-20\cdot 10^{-20}$J,
$k_A/t_0^2 \sim 6 \cdot 10^{-20}$J/nm${}^4$, $c_0 \sim -0.04$ nm${}^{-1}$. The
agreement is remarkable, given that our model is highly simplified and has not
been adjusted to reproduce the elastic properties of the bilayers. Our findings
indicate that the elastic material parameters of membranes are largely generic
and determined by global quantities such as the ratio of lipid area and squared
membrane thickness (which is the quantity that was ''adjusted'' to map our
model membranes on real lipid bilayers as discussed above).

Over all, we can conclude that the elastic theory gives an excellent
description of the fluctuations of tensionless fluid membranes.  We will now
consider the less well-studied case of membranes subject to a frame tension
$\Gamma_{\mbox{\tiny frame}}$.  As pointed out earlier, the ''tensionless''
case is a rather academic one since most membranes in nature are slightly under
stress.  However, even such slightly stretched membranes can be considered
quasi-tensionless in the ''floppy'' regime\cite{FAP01} where the stretching
energy is small compared to the bending energy on the length scales ($l$) of
interest, $\Gamma_{\mbox{\tiny frame}} l^2 \ll k_c$. For larger tensions and/or
on larger length scales, $l > \sqrt{k_c/\Gamma_{\mbox{\tiny frame}}}$, the
fluctuations are dominated by the tension and follow a $q^{-2}$-behavior. If
the stretching energy becomes large on molecular length scales ($l_0$),
$\Gamma_{\mbox{\tiny frame}} l_0^2 > k_B T$, the tension may also induce
structural changes in the membrane\cite{NWN10}. Such strongly stretched
membranes are only metastable and will eventually rupture, but they can exist
on short time scales, e.g., under the influence of transverse ultrasonic
pulses, and the effect of such pulses on membranes is of considerable medical
interest\cite{Mitragotri05}. Therefore, a number of simulation studies have
also been devoted to membranes under tension\cite{FZP95,FP96,FP99, Imparato06,
SOB06,KKY06,KKY08,Stecki08,NWN11} or compressed membranes\cite{Otter05}.

\begin{figure}
  \centering
  \includegraphics[width=0.6\textwidth]{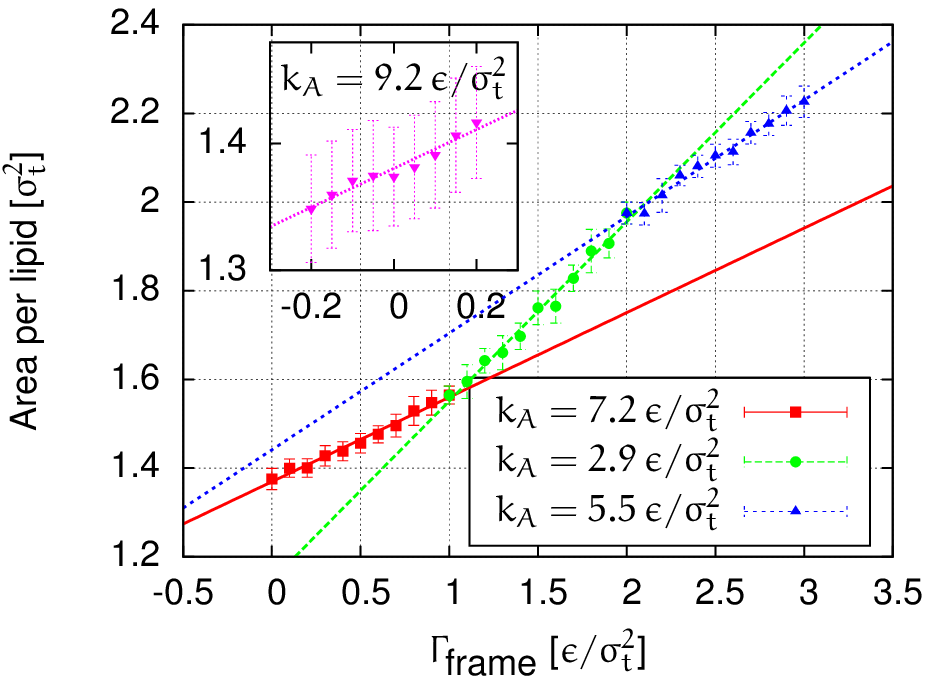}\\
  \includegraphics[width=1.0\textwidth]{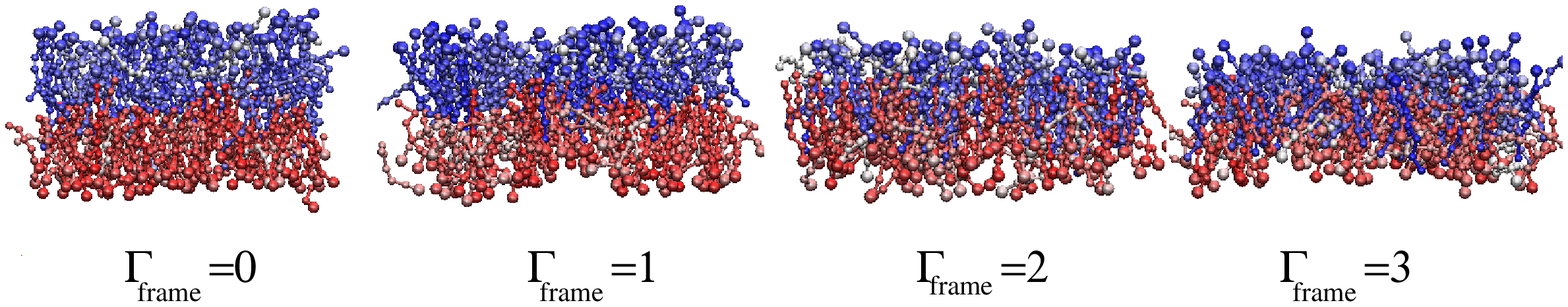}\\
  \caption{Top: area per lipid vs. tension in the fluid phase 
  ($T=1.3 \epsilon/k_B$), together with linear fits to the regimes
  $\Gamma_{\mbox{\tiny frame }} = 0-1 \epsilon/\sigma_t^2$, 
  $\Gamma_{\mbox{\tiny frame}} = 1-2 \epsilon/\sigma_t^2$, 
  and $\Gamma = 3-4 \epsilon/\sigma_t^2$. Inset focusses on the
  quasi-tensionless regime. 
  Bottom: Slices through bilayer configurations in the fluid
  phase ($T=1.3 \epsilon/k_B$) at different tensions as indicated
  (in units of $\epsilon/\sigma_t^2$). Chains pointing
  upward (from head to tail) are shown in red, chains pointing
  downward are shown in red. The size of the beads are not to scale.
  From Refs.\ \protect\onlinecite{Neder_Thesis,NWN10}.
  }
  \label{fig:conf_tension}
\end{figure}

In our simulation model, the strongly-stretched regime is reached at frame
tensions around $\Gamma_{\mbox{\tiny frame}} \sim 1 \epsilon/\sigma_t^2$. Fig.\
\ref{fig:conf_tension} shows the lipid area as a function of applied tension.
One can clearly distinguish three regimes with different compressibilities, two
stiffer ones at low and high $\Gamma_{\mbox{\tiny frame}}$
($\Gamma_{\mbox{\tiny frame}} < 1 \epsilon/\sigma_t^2$ and $\Gamma_{\mbox{\tiny
frame}} > 2 \epsilon/\sigma_t^2$) and a softer regime in between. A similar
softening at reduced pression one (in natural model units) was also 
observed in a different membrane model by Goetz and Lipowsky \cite{GL98}.  The
softening is associated with structural rearrangements within the membrane such
that the upper and lower monolayers become increasingly interdigitated (see
snapshots in Fig.\ \ref{fig:conf_tension}). At very high tensions, it becomes
difficult to distinguish between the monolayers; the membrane rather has the
structure of a single, disordered, interdigitated sheet.  Nevertheless, the
general considerations of Section \ref{sec:general} should still hold. Nowhere
did they depend on the fact that one deals with well-separated monolayers. The
same considerations can be made if $z_1$ and $z_2$ simply parametrize the two
outer surfaces of the membrane. Therefore, Eqs.\ (\ref{eq:elastic_height}) and
(\ref{eq:elastic_width}) should still apply, with the parameter $\Gamma$ being
the frame tension $\Gamma_{\mbox{\tiny frame}}$. 

\begin{figure}
  \centering
  \includegraphics[width=1.0\textwidth]{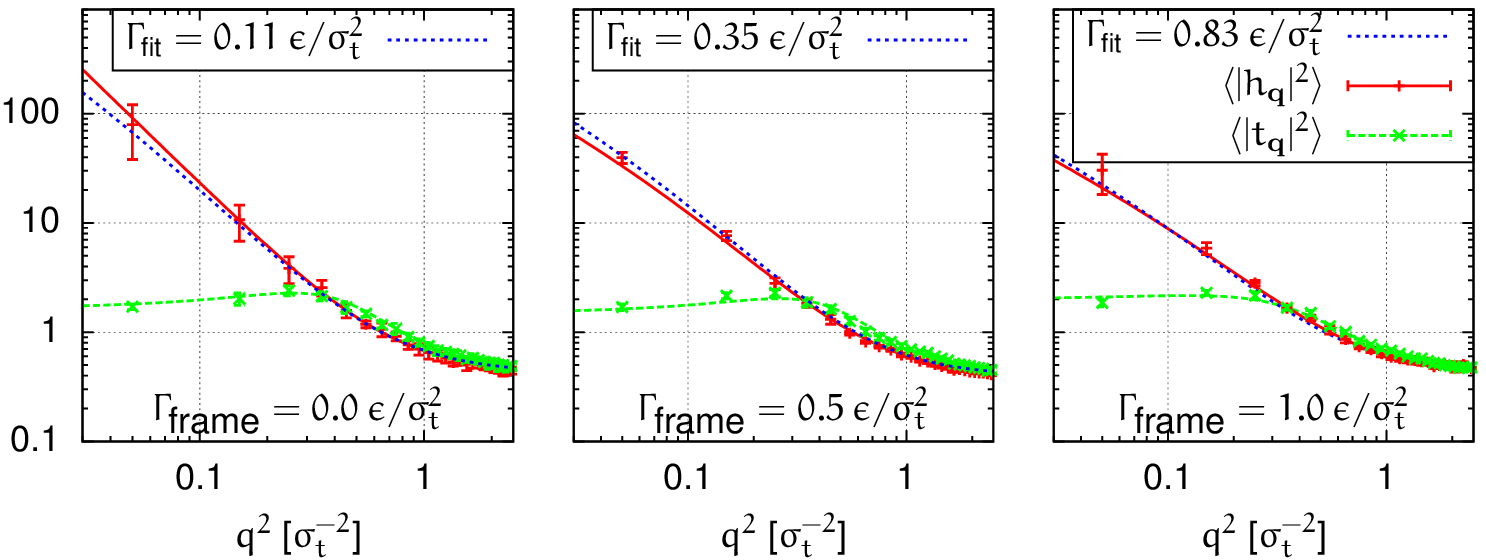}\\
  \includegraphics[width=1.0\textwidth]{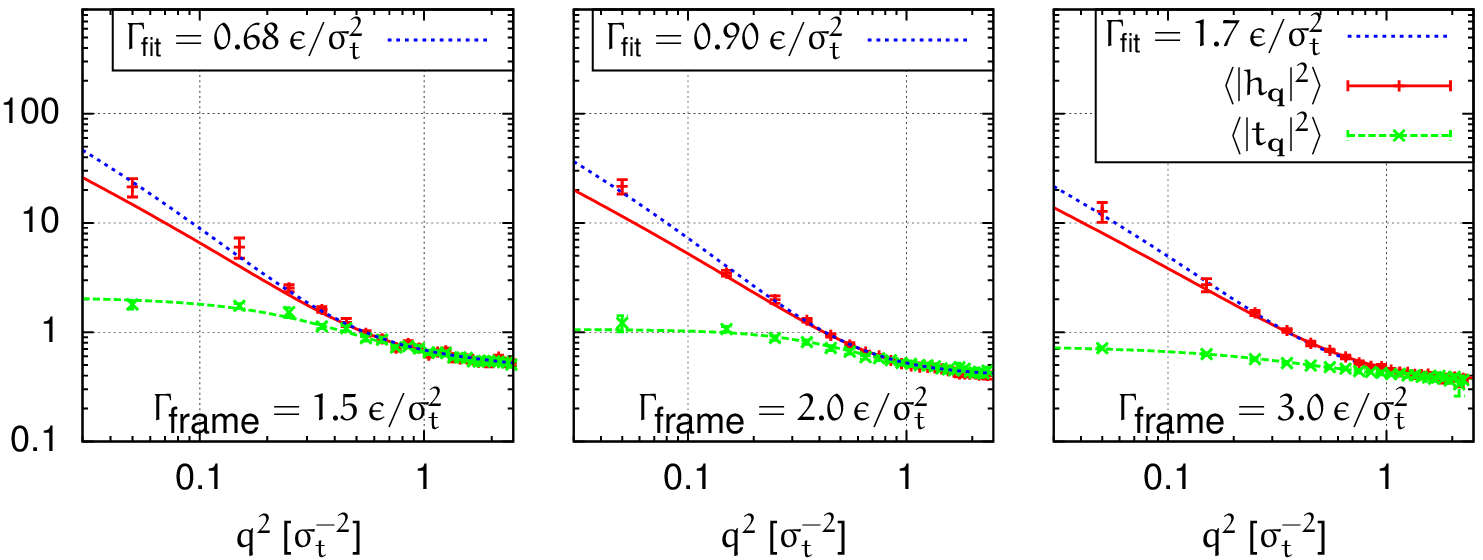}\\
  \caption{Fourier spectrum of the height (plus) and thickness (cross)
fluctuations of membranes under tension for different values of applied
tension. Solid lines show fits of the height spectrum to the elastic theory,
Eq.\ (\ref{eq:elastic_height}) and (\ref{eq:elastic_width}), if $\Gamma$ is
fixed at the value of the applied frame tension, $\Gamma = \Gamma_{\mbox{\tiny
frame}}$; dotted line give fits if $\Gamma$ is fitted as well.  Dashed lines
show fits to the thickness spectrum, which were almost identical in both cases.
From Ref.\ \protect\onlinecite{Neder_Thesis}, see also Ref.\
\protect\onlinecite{NWN10}.  }
  \label{fig:fluc_tension}
\end{figure}

In the simulations, however, the elastic fit is only satisfactory at low to
moderate tensions. In the strongly stretched regime at $\Gamma_{\mbox{\tiny
frame}} > 1 \epsilon/\sigma_t^2$, the fits fail to reproduce the low-$q$
behavior of the height fluctuations; they consistently underestimate the height
of the long-wavelength fluctuations.  Good fits can be obtained if the
parameter $\Gamma$ is taken to be a fit parameter, but the fitted values
$\Gamma_{\mbox{\tiny fit}}$ are then reduced by almost a factor of two,
compared to the true frame tension. Thus membranes under strong tension are 
distinctly softer than predicted by theory on large length scales.

At this point, we recall the discussion of the relation between the frame
tension $\Gamma_{\mbox{\tiny frame}}$ and the fluctuation tension
$\Gamma_{\mbox{\tiny fluc}}$ in Section \ref{sec:general}. There, we had
provided strong evidence for $\Gamma_{\mbox{\tiny frame}}= \Gamma_{\mbox{\tiny
fluc}}$ from simulations of simple structureless one-dimensional membranes. If
fluctuations do not renormalize $\Gamma_{\mbox{\tiny fluc}}$ in one dimension,
one would not expect them to do so in two dimensions, since the effect
of fluctuations is usually smaller in higher dimensions. 

Nevertheless, the effective fluctuation tension, $\Gamma_{\mbox{\tiny fluc}}
\equiv \Gamma_{\mbox{\tiny fit}}$, in our molecular simulations is clearly
smaller than the frame tension, $\Gamma_{\mbox{\tiny frame}}$ at high tensions
$\Gamma_{\mbox{\tiny frame}}$.  The main difference to the simple model
discussed in Section \ref{sec:general} is that our molecular membranes have internal structure.  As
mentioned earlier, Imparato et al and Stecki have also reported
discrepancies between $\Gamma_{\mbox{\tiny frame}}$ and $\Gamma_{\mbox{\tiny
fluc}}$ in molecular membrane simulations
\cite{Imparato05,Imparato06,Stecki08}. In these studies, $\Gamma_{\mbox{\tiny
fluc}}$ was found to be larger than $\Gamma_{\mbox{\tiny frame}}$, which is
opposite to our observation. Furthermore, they report discrepancies
already in the zero tension regime, where we find $\Gamma_{\mbox{\tiny fluc}}
\approx \Gamma_{\mbox{\tiny fluc}} \approx 0$ in agreement with most other
simulation studies\cite{GGL99,LE00,MM01,CD05,WF05,BB06,WBS09,BBS11} and also
with the theoretical arguments of Cai et al \cite{CLN94} and Farago
\cite{FP04,Farago11}. We note that the latter arguments are general enough that they
should also apply to tensionless membranes with internal structure (subject to
question marks regarding the thermodynamic limit as discussed in Section
\ref{sec:general}).  However, they {\em cannot} be applied to strongly stretched
membranes, since the stretching breaks rotational symmetry.

So far, we have no explanation for our findings. We can only speculate that it
must be a nonlinear effect, related to higher order terms in the Hamiltonian.
For example, one can argue that there should be a higher order coupling between
thickness and height fluctuations of the form ${\cal H}' \sim -
\frac{\Gamma_{\mbox{\tiny frame}}}{t_0} \int {\rm d}x {\rm d}y \: u \: (\nabla
H)^2$, which might renormalize $\Gamma$ to a value $\Gamma_{\mbox{\tiny fluc}}$ that
differs from $\Gamma_{\mbox{\tiny frame}}$ at high tensions \cite{NWN10}.  
It is remarkable that we observe the deviations in a tension regime where
the tension induces structural rearrangements in the membrane. 
The softening might be a signature of a nearby phase transition, or of
an instability with respect to rupture. However, no sign of actual rupture events
were observed in the membranes on the time scales of the simulations.

\subsection{Gel phase}

\label{sec:gel}

\begin{figure}
  \centering
  \includegraphics[width=0.6\textwidth]{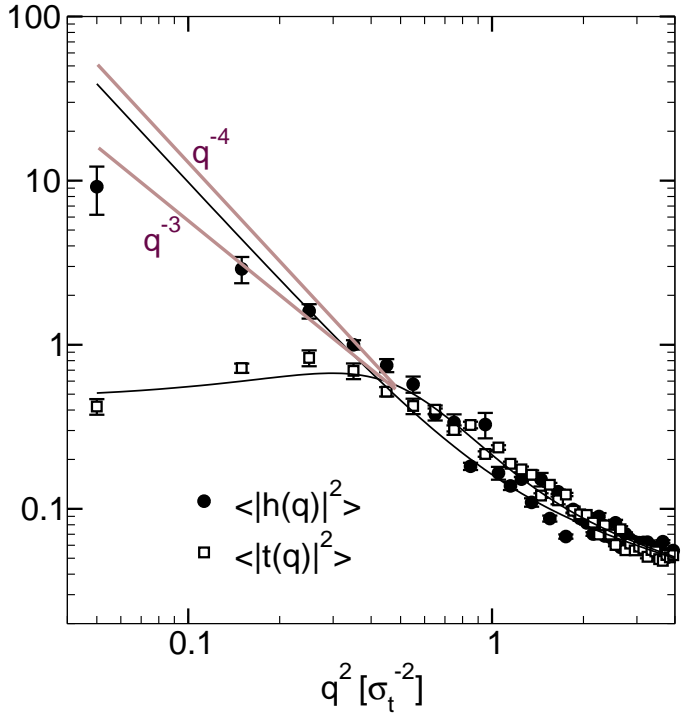}\\
  \caption{Radially averaged Fourier spectrum of the height (full circles)
  and thickness (open squares) fluctuations of membranes in the gel state 
  ($T=1 \epsilon/k_B$). Black solid lines show the best fit to Eqs.\
  (\protect\ref{eq:elastic_height}) and (\protect\ref{eq:elastic_width}).
  Grey solid line indicates the slopes corresponding to a
  $q^{-4}$-behavior and a $q^{-3}$-behavior. 
  After Ref.\ \protect\onlinecite{WS10}.
  }
  \label{fig:fluc_gel}
\end{figure}

The situation becomes even worse in the gel phase. Fig.\ \ref{fig:fluc_gel}
shows the radially averaged thickness and height fluctuation spectra for
tensionless membranes in the $L_{\beta'}$ phase\cite{WS10} ($T=1
\epsilon/k_B$).  At first sight, the qualitative features seem similar to those
in the fluid phase, even though the amplitudes of the fluctuations are much
smaller. As in the fluid phase, the height spectrum diverges at low $q$, the
thickness spectrum features a soft peristaltic mode at wavelengths around $10
\sigma_t$, and both spectra are nearly identical at high $q$. The fit to the
elastic theory, Eqs.\ (\ref{eq:elastic_height}) and (\ref{eq:elastic_width}),
is not as good as in the fluid state. Most notably, long-wavelength
fluctuations are suppressed, compared to the theoretical prediction, and one
never encounters a bending regime with $\langle | h(q)|^2 \rangle \sim q^{-4}$.
This is not surprising, given that one expects the shearing modes in gel
membranes to renormalize the bending rigidity at small $q$, as discussed in the
introduction. The in-plane shear modes introduce effective interactions between
height fluctuation modes, which lead to a stiffening on large length scales
\cite{NP87}. However, even the strongest possible renormalization $k_c(q)
\propto q$, corresponding to quasi-crystalline membranes\cite{NP87}, does not
fully account for the discrepancy. It would result in a $q^{-3}$ divergence of
the height fluctuations at small $q$, which is still too steep. 

\begin{figure}
  \centering
  \includegraphics[width=0.6\textwidth]{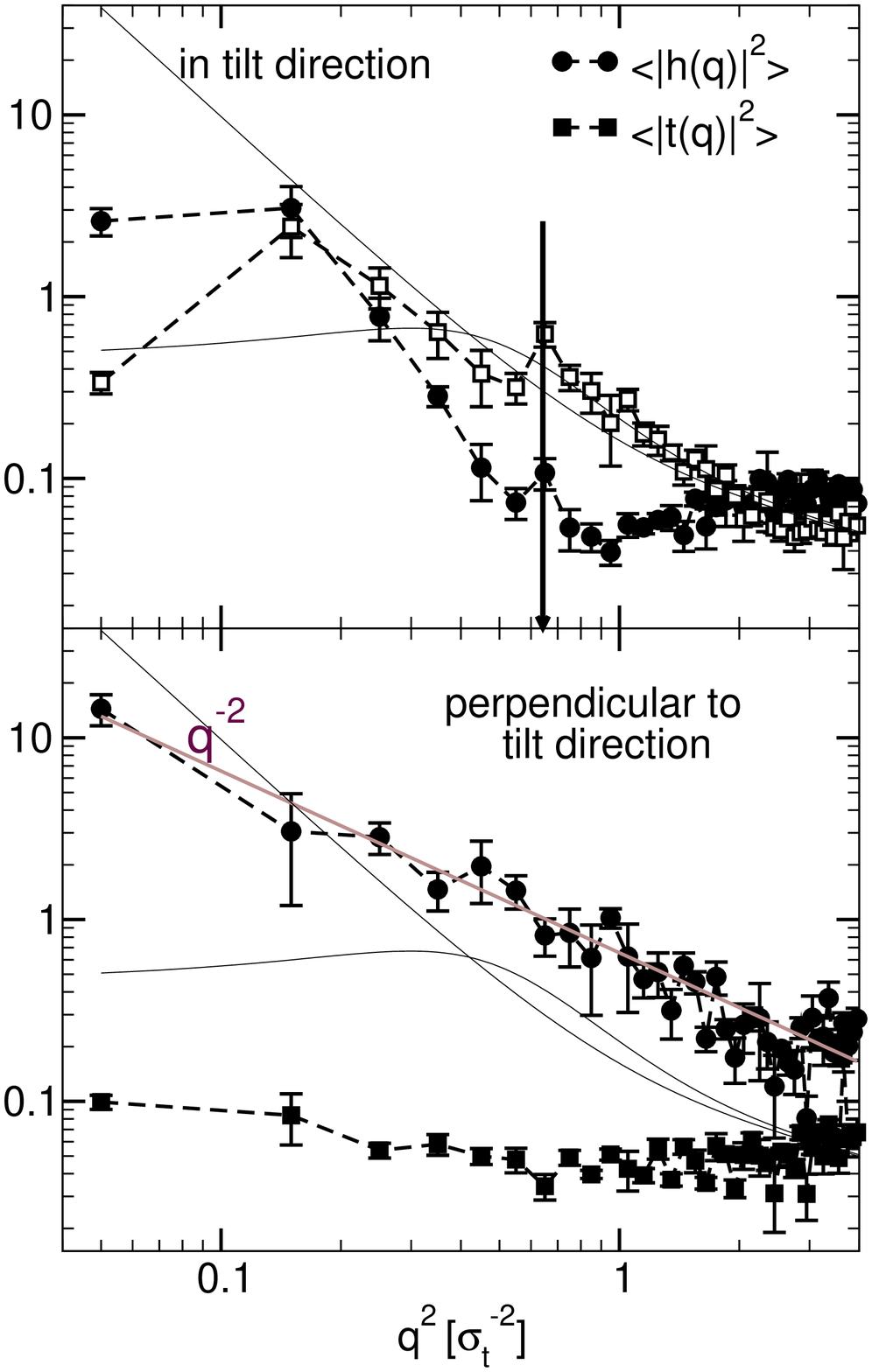}\\
  \caption{
  Fourier spectrum of the height (full circles) and thickness (open squares) 
  fluctuations of the same membranes of Fig.\ \protect\ref{fig:fluc_gel} 
  for wavevectors parallel (top) and perpendicular (bottom) to the tilt direction.
  The thin solid lines reproduce the thin solid lines from
  Fig.~\protect\ref{fig:fluc_gel} (the fit of the radially averaged
  spectra to the elastic theory) for comparison. Vertical arrows
  in upper panel mark the positions of peaks in both the height and 
  thickness fluctuation spectrum. Thick grey line in lower panel
  indicates $q^{-2}$-slope. After Ref.\ \protect\onlinecite{WS10}.
  }
  \label{fig:fluc_gel_xy}
\end{figure}

Despite these quantitative discrepancies, Fig.\ \ref{fig:fluc_gel} suggests
that the elastic theory describes the fluctuations of gel membranes reasonably
well except for the large-wavelength height fluctuations. Unfortunately, the
radial average shown in Fig.\ \ref{fig:fluc_gel} hides the true extent of the
problem. It becomes apparent when one inspects the fluctuations in tilt
direction and perpendicular to the tilt direction separately, as shown in Fig.\
\ref{fig:fluc_gel_xy}. The fluctuation spectra in these directions differ
strongly from the radially averaged spectra, and they are not at all compatible
with the elastic theory.  The height fluctuations are mostly suppressed in the
direction of tilt (with the exception of a broad peak around $q^2 \sim 0.15
\sigma_t^2$ to be discussed below), whereas they are relatively strong in the
perpendicular direction. The thickness fluctuations show the opposite trend.
Their spectrum mostly matches the radially averaged thickness spectrum in the
direction of tilt, and it is almost entirely suppressed in the perpendicular
direction. 

Two features of these directionally resolved spectra are particularly
remarkable.  First, both the thickness and the height fluctuation spectra in
the direction of tilt feature distinct maxima around $q^2 \sim 0.15/\sigma_t^2$
and $q^2 \sim 0.65/\sigma_t^2$, corresponding to soft modes with wavelength
$\lambda \sim 8 \sigma_t$ and $2 \lambda \sim 16 \sigma_t$. The wavelength
$2\lambda$ coincides with the period of the ripple phase\cite{LS07}. Thus the
fluctuations in tilt direction carry the signature of the modulated phase which
intrudes between the liquid and the gel state.  

Second, the amplitudes of the height fluctuations in the perpendicular
direction follow a $q^{-2}$ behavior over the whole $q$-range, suggesting that
they are controlled by an interfacial tension rather than a bending rigidity.
Such a behavior can be rationalized if one assumes that the shape fluctuations
of the membrane are governed by ''out-of-layer'' fluctuations of the lipids
(up- and down displacements without splaying) rather than bending modes which
involve lipid splay\cite{WS10}.  One would expect the out-of-layer fluctuations
to eventually give way to bending-dominated fluctuations on very large length
scales, but this is not observed in our simulations. 

In sum, the elastic theory does not describe adequately the fluctuations in the
gel phase, even though a naive inspection of the radially averaged fluctuation
spectra suggests otherwise. Due to the ordered anisotropic structure of the gel
membrane, the fluctuation spectra are highly anisotropic and they carry the
signature of the ripple phase.  On very large length scales, the fluctuation
spectra will presumably adopt an asymptotic behavior which is compatible with
the theoretical expectations outlined in the introduction. However, our
findings suggest that the microscopic unit in such a description is not the
molecular length scale, but rather a multiple of the period of the ripple
phase, which is of order $10 {\rm nm}$. Moreover, the anisotropy imprinted on
the membrane shapes by the tilt order of the lipids may persist on length
scales up to micrometers\cite{LDW03}, and only on even larger length scales can
then the membrane be treated as an isotropic polycrystalline sheet with
multiple tilt directions. Based on these considerations, we expect that the
asymptotic behavior will only be reached on length scales much larger than
hundreds of nanometers, if not micrometers. Our simulation systems, which had
linear dimensions of around $\sim 35 {\rm nm}$, where much too small to capture
this limit. On these smaller length scales, the elastic behavior results from a
complex interplay of contributions which cannot easily be told apart, and
suitable continuum theories are still to be developed.  Such continuum models
might also contribute to a better understanding of the mechanisms driving the
formation of the ripple phase.  

\section{Conclusions}

The analysis of membrane fluctuations provides an important link between
molecular models and elastic continuum models.  Despite the fact that
fluctuation analyses have a long tradition, there are still amazingly many open
questions.  Membrane undulations can be described and understood at a
quantitative level for fluid and tensionless (or floppy) membranes. In this
case, the fluctuation analysis can be used to extract elastic parameters of the
membranes, which can then be used to bridge between different representations
of membranes in the context of multiscale modeling approaches.  In all other
cases -- stretched membranes under tension, membranes in phases that are other
than fluid -- the connection between molecular model and an appropriate
continuum theory does not yet succeed satisfactorily.  Much work still 
remains to be done.

\section{Acknowledgments}

The author thanks Frank Brown, Stefan Dolezel, J\"org Neder, Peter Nielaba, 
Olaf Lenz, Sebastian Meinhardt, and Beate West for enjoyable collaborations.
The work described in this article was funded by the German Science Foundation
(DFG) within the SFB 613 and SFB 625. The simulations were carried out 
at the Paderborn center for parallel computing (PC2) and the 
John von Neumann Institute for Computing in J\"ulich. 

%%%%%%%%%%%%%%%%%%%%%%%%%%%%%%%%%%%%%%%%%%%%%%%%%%%%%%%%%%%%%%%%%%%%%
%% The appropriate \bibliographystyle and \bibliography commands
%% should be placed here.
%%%%%%%%%%%%%%%%%%%%%%%%%%%%%%%%%%%%%%%%%%%%%%%%%%%%%%%%%%%%%%%%%%%%%
\bibliographystyle{apsrev}
\bibliography{paper}

\end{document}